# GRAPHENE-LIKE METALLIC-ON-SILICON FIELD EFFECT TRANSISTOR

M. Dragoman[1], G. Konstantinidis[2], K. Tsagaraki[2], T. Kostopoulos[2], D. Dragoman[3*], D. Neculoiu[4],

[1] National Institute for Research and Development in Microtechnology (IMT), P.O. Box 38-160, 023573 Bucharest, Romania

[2] Foundation for Research & Technology Hellas (FORTH) P.O. BOX 1527, Vassilika Vouton, Heraklion 711 10, Crete, Hellas

[3] Univ. Bucharest, Physics Dept., P.O. Box MG-11, 077125 Bucharest, Romania

[4] Politehnica University of Bucharest, Electronics Dept., 1-3 Iuliu Maniu Av., 061071 Bucharest, Romania

## Abstract

This paper presents a field effect transistor with a channel consisting of a two-dimensional electron gas located at the interface between an ultrathin metallic film of Ni and a $p$-type Si(111) substrate. We have demonstrated that the two-dimensional electron gas channel is modulated by the gate voltage. The dependence of the drain current on the drain voltage has no saturation region, similar to a field effect transistor based on graphene. However, the transport in this transistor is not ambipolar, as in graphene, but unipolar.

---

*Corresponding author: danieladragoman@yahoo.com

Graphene is a graphite monolayer with a thickness of only 0.34 nm, formed of carbon atoms in a hybridization state of sp$^2$, so that each atom is covalently bonded to three others. Graphene is also a two dimensional (2D) crystal and a natural 2D gas of charged particles. The ballistic transport of carriers in graphene takes place at room temperature over a distance of 0.4 µm and reaches an intrinsic mobility[1] of 44 000 cm$^2$V$^{-1}$s$^{-1}$. However, when graphene is deposited with a hexagonal boron nitride substrate[2] matching the graphene lattice, mobilities higher than 100 000 cm$^2$/s and mean-free carrier paths of 1 µm are measured at room temperature.

Due to these electrical properties, that are superior by orders of magnitude to those of semiconductors and semiconductor heterostructures, graphene is seen as a promising material for ultrafast nanoelectronic devices such as transistors with cutoff frequencies beyond 100 GHz, and optical devices, for example photodetectors; it could also be used as a transparent electrode for photovoltaic applications.[3,4] However, the growth of graphene at the wafer level is still a technological challenge because the electrical properties are seriously downgraded due to various defects. Even today the best electrical properties of graphene are obtained from graphene flakes mechanically exfoliated with an adhesive tape from graphite[5].

Therefore, it is very tempting to seek alternative ways to attain high mobility in new materials, with values comparable to those in graphene. In addition, the devices based on these new materials should be manufacturable at the wafer scale, allowing standard clean room processes. Along this path, a recent experiment demonstrated that very small carrier effective masses could be obtained at a metal/$n$-S(111) interface, where the metal is of the order of one (or few) monolayer(s) thick[6]. Reference 6 demonstrates that in the ultrathin metal film/$n$-Si(111) nanostructure, the surface states favor the creation of a quantum well in the Si inversion layer and reduces drastically the effective mass of charge carriers in Si due to repulsive forces. As a consequence, Si(111) displays an almost linear dispersion relation and a similar carrier effective mass as in graphene, of only $m_{eff} = 0.0075\ m_0$, where $m_0$ is the electron rest mass; this value is 20 times smaller than in bulk Si. Moreover, the surface states adjacent to the induced quantum well are localized states, in which electrons are totally confined, and play an analogous role to the gate dielectric in a normal FET (field-effect transistor). The selection of Si(111) as a substrate is not accidental. It has a honeycomb lattice, similar to that of graphene, and calculations imply that the dispersion relation could be linear close to the K point, with the (massless) Dirac fermions moving with a velocity[7] of $10^4$ m/s.

The aim of this paper is to corroborate the predictions made in Ref. 6 by fabricating a FET based on a graphene-like nanomaterial consisting of *p*-Si(111) covered with an ultrathin metallic layer, and thus to pave the way towards Si-based ultrafast electronics. We expect this FET, which is readily reproducible on a Si(111) wafer using standard micro/nanotechnologies, to have similar properties to graphene FETs.

It is not trivial to deposit a metal on Si(111) with a thickness of 1 nm with UHV electron gun evaporation. Initially, we investigated the quality of our metallization in terms of roughness and full surface coverage. For this purpose, we evaporated very thin layers of Ir, Pt, Ni, Al, Au and Ag. Through AFM investigations we concluded that the minimum metal thickness needed for full surface coverage was 1-2 nm. The results in terms of roughness are summarized in Table 1.

For the first experiment we decided to use Ni contacts since Ir and Pt deposition requires substrate heating to secure proper adhesion, which is often incompatible with conventional photoresist-based metal lift-off techniques. We employed an already existing set of lithography masks designed for the fabrication of DC and RF transistors on GaAs-based FET and HEMT structures. The gate length was $L = 2$ μm and the gate width was $W = 180$ μm, while the source S and drain D regions had dimensions of 80 × 320 μm$^2$ and 80 × 80 μm$^2$, respectively. The key issue was the need for the S and D contacts to be self aligned with the gate. Thus, we defined the gates by contact UV lithography on a *p*-Si(111) substrate. Then, we deposited 2 nm of Ni and another 500 nm of SiO$_2$ on top with e-gun evaporation. The final pattern was defined with lift-off. The next step was the use of the image reversal photoresist process in order to define the "negative" image of the MESA mask. For this purpose, a SF$_6$-based shallow reactive ion etching (RIE) of 300 nm was first performed, followed by the evaporation of 280 nm Mo as S and D contacts. Lift-off was again employed to define the patterns. Care was taken for the S and D metallization to not exceed the etched depth to avoid shortening with ultrathin gate. However, with this fabrication sequence, S and D metals still exist on the SiO$_2$ after the lift-off. Thus, this excess metal layer was removed following the dissolution of the SiO$_2$ in BHF. Then, the whole sample was covered with 200 nm of PECVD grown SiN. Finally, pad openings were defined and etched in the SiN with RIE and the metallic pads were fabricated with the deposition of 200 nm Au and metal lift-off. The scanning electron microscopy (SEM) photo of the transistor is displayed in Fig. 1a and a schematic cross-section of the device is depicted in Fig. 1b.

The channel of this FET is located at the interface of the very thin metallic film and *p*-Si(111). At this interface, an inversion layer appears in Si due to the very strong bending of the energy bands of Si(111) as a result of the deposited ultrathin film and a quantum well is formed in the conduction band of Si. The Fermi level is shifted inside this *n*-type quantum well and forms a 2D electron gas (2DEG) that is further controlled by an applied gate voltage. The reduction in the effective mass of charge carriers in Si due to the surface states in the ultrathin Ni film is expected to manifest itself in a high mobility value of the FET. Electrical characterization of the FET was performed using a Keithley 4200 station. The DC characteristics were measured directly on wafer for tens of similar transistors which have displayed very reproducible results. The drain current ($I_D$) versus drain voltage ($V_D$) characteristic is displayed in Fig. 2 at various gate voltages ($V_G$), which increase from 0 V, to 0.642 V, 1.07 V and 1.5 V. For negative drain voltages, the gate did not influence the drain current value. We observe that the transport is dominated by electrons, indicating the formation of an inversion layer, as predicted by Ref. 6; in our case the inversion layer is *n*-type because the substrate is *p*-type. There is no ambipolar transport, as in graphene, because the *p*-Si(111) substrate has a finite bandgap. However, as in graphene, the drain current is not saturated, meaning that the carrier concentration depends on the gate voltage. More precisely, in a 2DEG channel with $M$ transverse modes, assuming that the kinetic energy of charge carriers is determined by the gate voltage $V_g$, the carrier density can be estimated as $n = (m_{eff}/\pi\hbar^2)MeV_g/\alpha$; this parameter is directly proportional to the gate voltage, as in graphene, although the proportionality constant is different. The number of modes can be found from the $M^2 = (eV_g/\alpha)/(\hbar^2\pi^2/2m_{eff}d_{2DEG})$, where $d_{2DEG}$ is the thickness of the 2DEG layer and $\alpha$ is a factor which describes the fraction of the voltage applied between the gate and the source electrodes that drops on the Ni gate metal. Since the largest part of the voltage is applied on the SiN protective layer, $\alpha$ takes high values.

Further, we have computed the mobility from the $I_D$-$V_D$ dependence using the formula $\mu = \dfrac{1}{en}\dfrac{\Delta I_D/W}{\Delta V_D/L}$. For $m_{eff} = 0.007m_0$ (see Ref. 6), $\alpha = 300$, and $d_{2DEG} = 10$ nm, it follows that $\mu = 12\,000$ cm$^2$/Vs. The coefficient $\alpha$ could be smaller for an optimized FET geometry. This high mobility value results from a significant decrease in the effective mass of electrons in the 2DEG, as predicted by Ref. 6, which leads to a relative small electron density. This mobility value is considerably larger than in other graphene-like FET transistors based on MoS$_2$ [8] and graphene nanoribbons [9] and is comparable to or even greater that in graphene FETs epitaxially grown on the carbon face of SiC [10] or in graphene FETs on $SiO_2$ substrates[11].

In Fig. 3 we have displayed the transistor's transconductance for three drain voltage values: 3 V (blue dashed line), 6 V (black solid line), and 9 V (red dotted line). This figure shows that the transconductance is almost constant with the gate voltage and has values comparable to those of many graphene FETs fabricated using various technologies [12].

In conclusion, we have obtained a Si-based FET transistor, which mimics the graphene FET transistor behavior, but can be fabricated at the wafer level. We expect that by employing a deep submicron gate self align process, the results will significantly enhanceThis research opens the way to ultrafast electronics based on charge carriers with small effective masses in Si.

*Acknowledgements* M.Dragoman states that this work was supported by a grant of Romanian National Authority for Scientific Research, CNCS-UEFISCDI, project number PN-II-ID-PCE-2011-3-0071. G. Konstantinidis acknowledges the "Graphene Center" of FORTH while all the authors the LEA SMARTMEMS

**Table I. The roughness measured with AFM for various metals with a 2-nm thickness**

| Material | RMS roughness (nm) in 0.5×0.5 μm² | RMS roughness (nm) in 1×1 μm² |
|---|---|---|
| Bare Si | 0.21 | 0.20 |
| Ir | 0.21 | 0.23 |
| Pt | 0.26 | 0.22 |
| Ni | 0.36 | 0.26 |
| Al | 0.43 | 0.40 |
| Au | 0.59 | 0.46 |
| Ag | 2.70 | 2.57 |

**Figure captions**

Fig.1 (a) The graphene-like FET transistor based on an ultrathin Ni layer deposited on a *p*-Si(111) substrate (b) the schematic cross-section of the FET

Fig. 2 The $I_D$-$V_D$ characteristics for a gate voltage of 0 V (black line), 0.642 V (red line), 1.07 V (green line) and 1.5 V (blue line).

Fig. 3 The trasconductance of the graphene-like FET for a drain voltage of 3 V (blue dashed line), 6 V (black solid line), and 9 V (red dotted line).

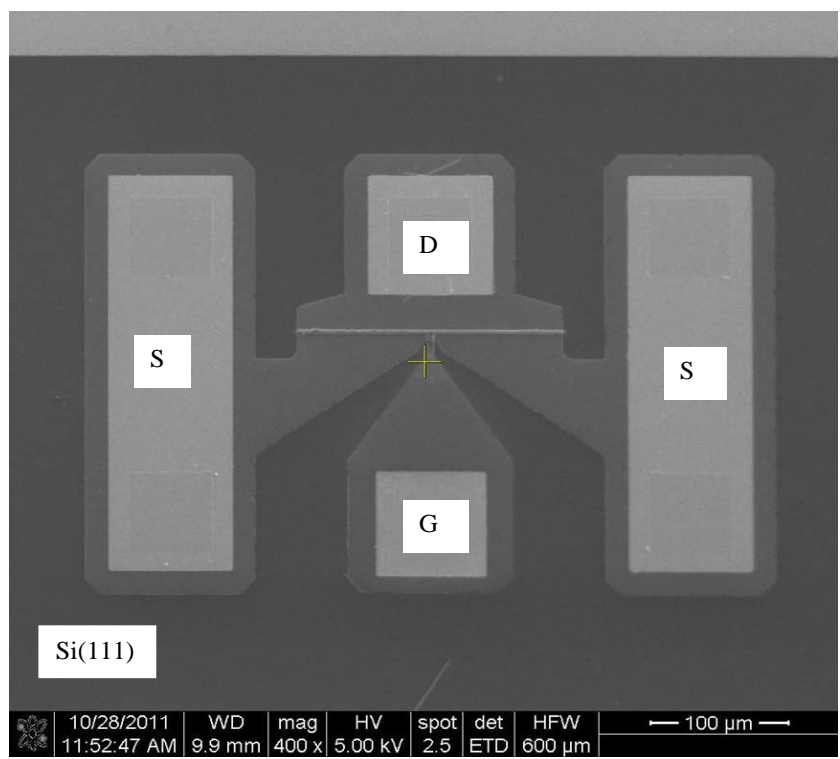

(a)

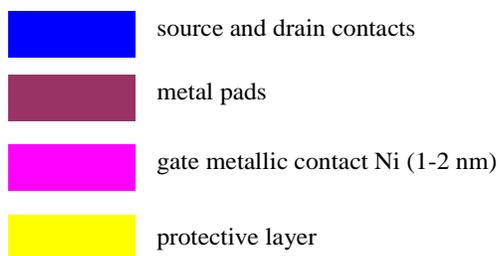

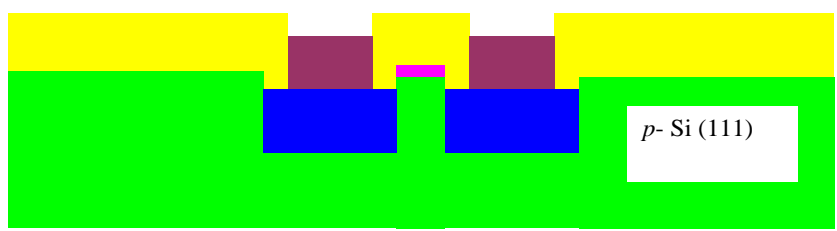

(b)

Fig.1

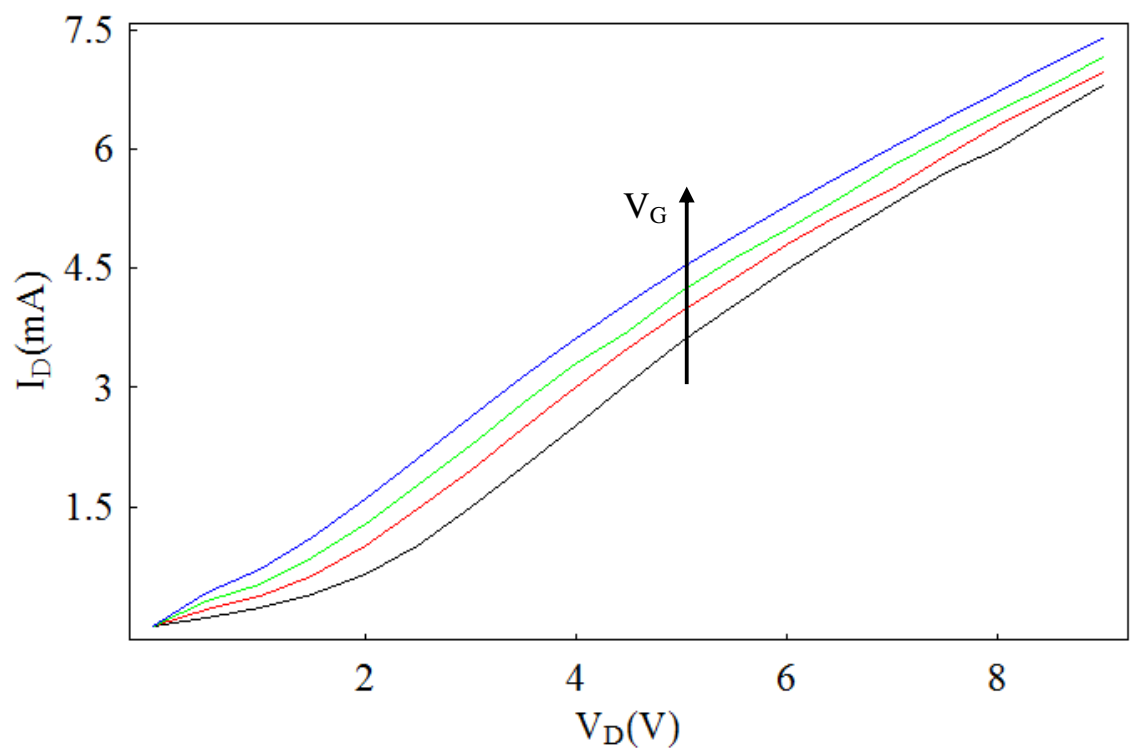

Fig. 2

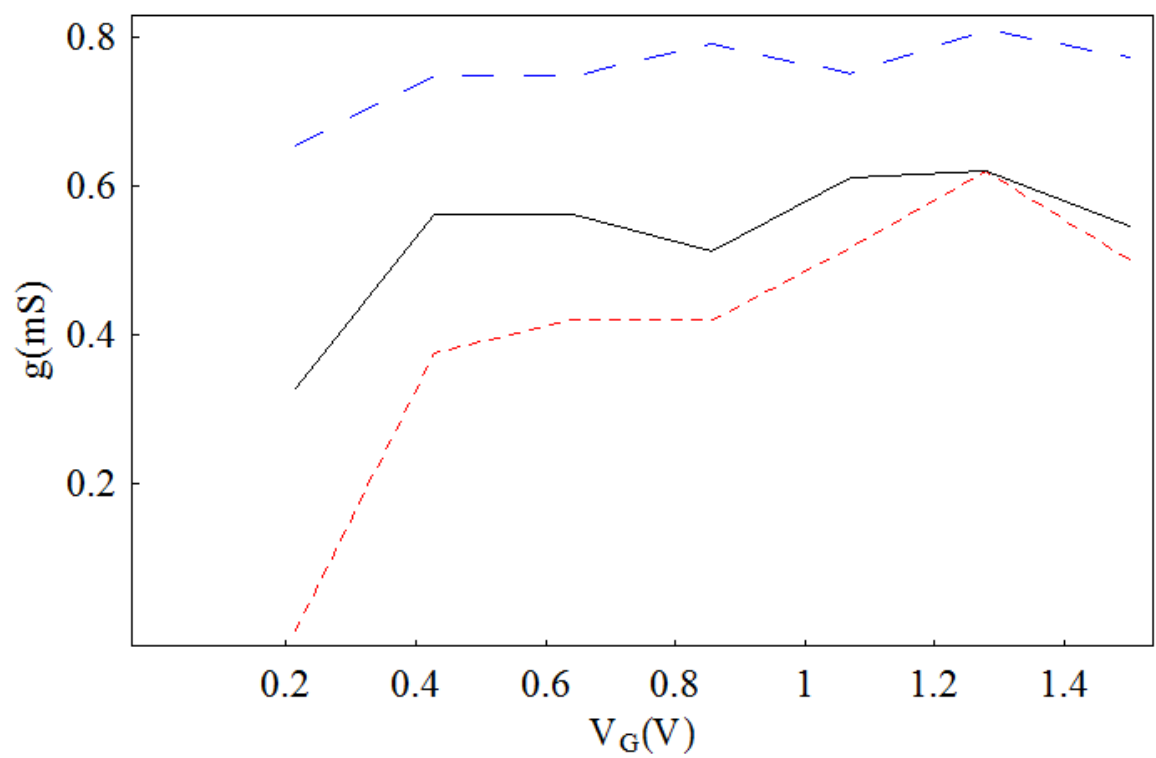

Fig.3